\documentclass[pre,showpacs,twocolumn,groupaddress]{revtex4}
\usepackage{epsfig,amsmath,amssymb,graphics,color,calc}

\newcommand{\be}{\begin{equation}}
\newcommand{\ee}{\end{equation}}
\newcommand{\ba}{\begin{eqnarray}}
\newcommand{\ea}{\end{eqnarray}}

\renewcommand{\phi}{\varphi}

\begin{document}

\title{The glass transition of dense fluids of 
hard and compressible spheres}

\author{Ludovic Berthier}
\affiliation{Laboratoire des Collo{\"\i}des, Verres
 et Nanomat{\'e}riaux, Universit{\'e} Montpellier II and CNRS,
34095 Montpellier, France}

\author{Thomas A. Witten}
\affiliation{James Franck Institute and Department of Physics,
The University of Chicago, 929 E. 57th Street, Chicago, Illinois 60637, USA}

\date{\today}
 
\begin{abstract}
We use computer simulations to study the glass transition of
dense fluids made of polydisperse, repulsive spheres. 
For hard particles, we vary the volume fraction, $\phi$, and use 
compressible particles to explore finite
temperatures, $T>0$. In the hard sphere limit, our dynamic data 
show evidence of an avoided mode-coupling singularity
near $\phi_{\rm MCT} \approx 0.592$, they are consistent
with a divergence of equilibrium relaxation times 
occuring at $\phi_0 \approx 0.635$, 
but they leave open the existence of a finite temperature
singularity for compressible spheres at volume fraction $\phi > \phi_0$.
Using direct measurements and a new scaling procedure, 
we estimate the equilibrium equation of state 
for the hard sphere metastable fluid up to $\phi_0$, 
where pressure remains finite, suggesting that $\phi_0$ corresponds
to an ideal glass transition.
We use non-equilibrium protocols to explore glassy states 
above $\phi_0$ and establish the existence of multiple equations 
of state for the unequilibrated glass of hard spheres, all diverging 
at different densities in the range $\phi \in [0.642, 0.664]$. 
Glassiness thus results in the existence of a continuum of densities 
where jamming transitions can occur.
\end{abstract}

\pacs{05.10.-a, 05.20.Jj, 64.70.Pf}


\maketitle

\section{Introduction}

Making sharp statements about the existence of 
genuine phase transitions underlying the observation
of non-equilibrium glassy states is a favourite game 
for workers dealing with disordered states of matter,
one which can easily prompt much controversy.
Experimentally, a notable  exception is the case of spin 
glasses for which the existence of a finite temperature 
phase transition is not controversial~\cite{young}, because the
location of the spin glass transition was consistently 
determined from theoretically motivated 
dynamic scaling relations~\cite{binder}, also reported 
for some frustrated magnets~\cite{gingras}. 
Despite the ubiquitous observation
of non-ergodic disordered states across 
condensed matter physics~\cite{houches,bookjamming}, the existence of 
genuine glassy phases is in fact very rarely 
established.

For many-body particle 
systems~\cite{reviewnature}, 
such as molecular and colloidal glasses, no such scaling predictions 
are available or experimentally accessible, 
and the location of glass transitions
is often deduced from a limited set of measurement
using uncontrolled extrapolations~\cite{reviewnature}. 
For molecular glasses, fitting the temperature evolution of the 
viscosity of a large number of materials 
even over more than 15 decades cannot qualitatively discriminate
theories based on the existence of a finite temperature 
singularity from those suggesting a divergence at zero-temperature 
only~\cite{fit1,fit2,fit3}.
Another well-studied instance where glassy behaviour is observed 
is the hard sphere system
at thermal equilibrium, which becomes highly viscous
when the packing 
fraction $\phi$ increases. Interestingly, this idealized
model system can be realized experimentally using colloidal
particles~\cite{pusey}. However, the determination of the 
location of a critical volume 
fraction where the equilibrium relaxation time diverges 
is plagued by uncertainties similar to the ones 
encountered in thermal glasses, since it relies 
on the extrapolation of a singularity from a single set of data
obtained at increasing $\phi$~\cite{vanmegen,chaikin,luca}. 

\begin{figure}[b]
\psfig{file=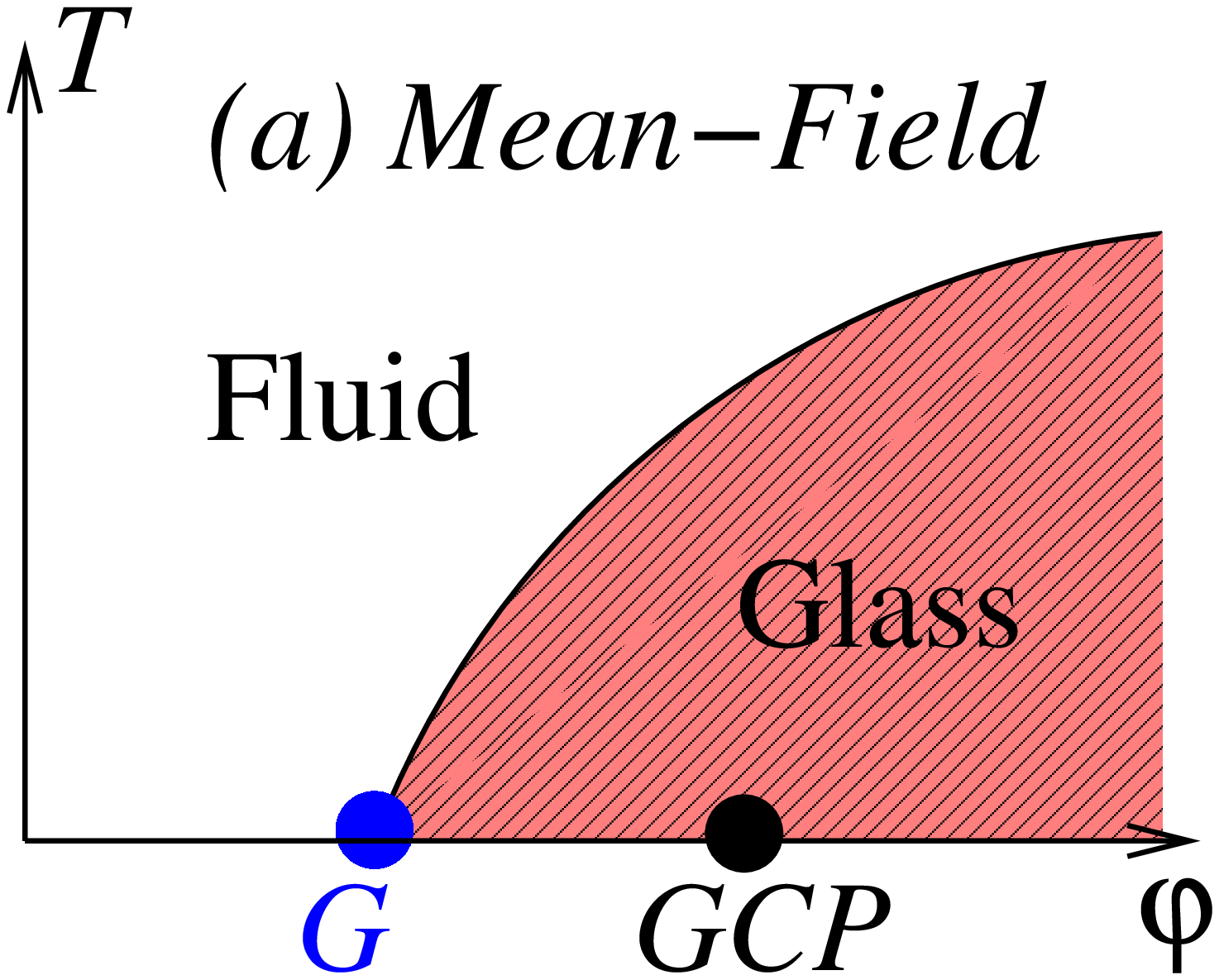,width=4.27cm}
\psfig{file=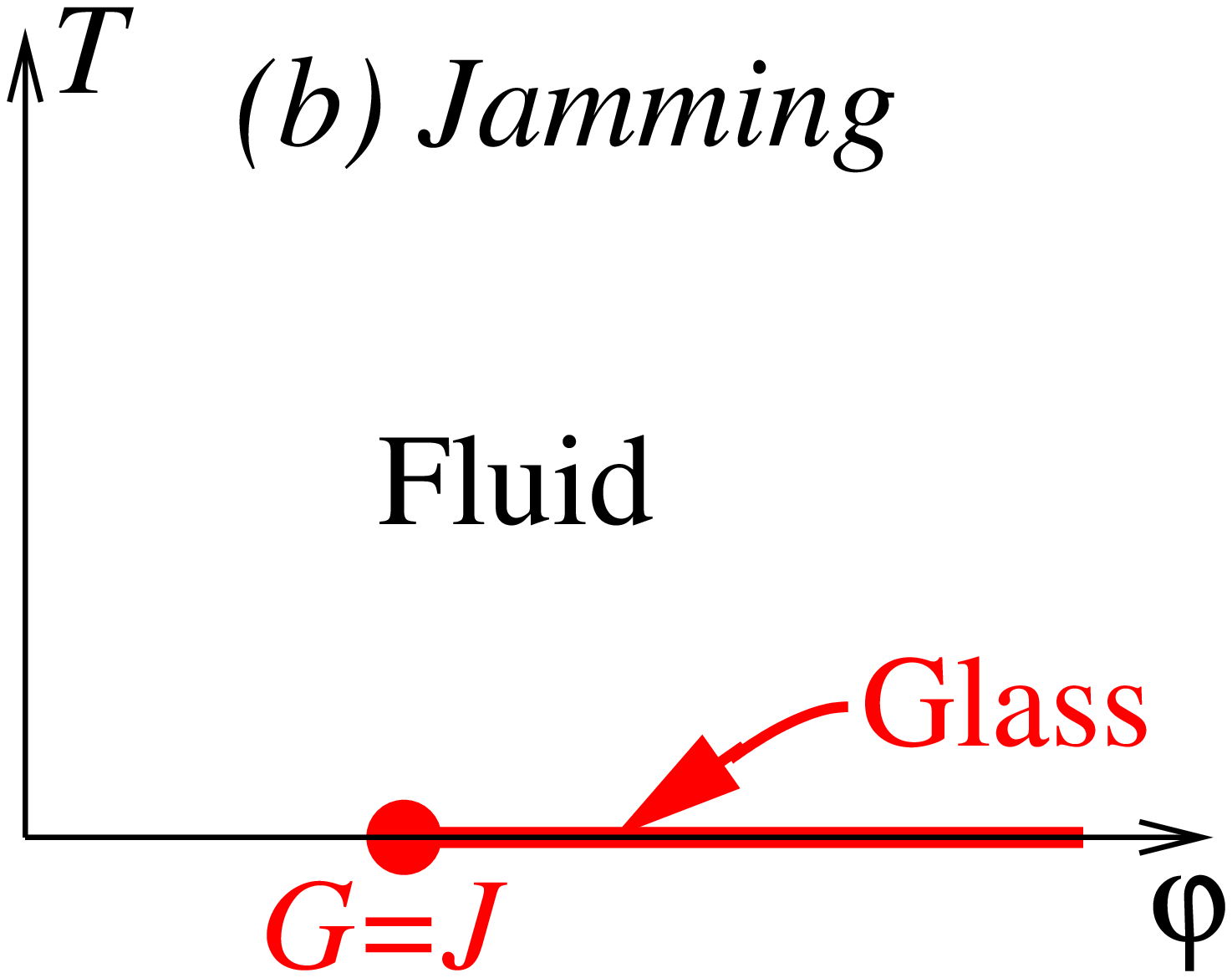,width=4.27cm}
\caption{\label{sketch} Sketch of two possible 
phase diagrams for harmonic spheres. (a) In mean-field replica 
calculations, 
an ideal glass transition 
occurs at finite temperature for 
$\phi$ above that of point $G$, the ideal glass transition 
point for hard spheres.
The pressure of the equilibrium 
glass diverges at larger density at
Glass Close Packing ($GCP$).
(b) No glass transition occurs before jamming 
at point $J$ at $T=0$ where the equilibrium pressure also 
diverges, with no glass transition 
occuring at finite temperature at larger density.
Alternative phase diagrams such as (a) with $G=GCP$, or 
(b) with $G \neq J$ are in principle possible.}
\end{figure}

In a recent article~\cite{tom}, we studied the glassy dynamics 
of a model of soft repulsive particles where the glass transition
occurs when either the particle volume fraction, $\phi$, or 
the temperature, $T$, are varied, see Fig.~\ref{sketch}.
We have discovered the existence of 
activated dynamic scaling in the whole $(T,\phi)$ 
plane constraining the actual functional form of the divergence, 
and allowing a precise location of the singularity.
In particular, we found that in the limit $T \to 0$, where 
the particles become 
infinitely hard, a dynamic singularity seems to occur at a 
well-defined critical packing fraction, $\phi_0$. We called 
`point $G$' the location of this special glass point
in the phase diagram shown in Fig.~\ref{sketch}.

Our aim in the present work is 
to study more precisely the nature 
of point $G$, extending our studies to thermodynamic observables, 
most notably equations of state, and 
to explore how point $G$ fits within existing theoretical frameworks.
By simultaneously studying hard and compressible particles, we are 
again able to put severe constraints on the nature of the dynamic divergence 
at point $G$. The picture which is most consistent with our data
relies on the existence of a hard sphere glass phase above $\phi_0$. 
It remains possible that the ideal glass transition reported here   
is eventually avoided when a much larger set of data becomes available, but 
we show that theories where a transition is absent describe the data
rather poorly.

We first introduce the theoretical background guiding our 
analysis (Sec.~\ref{theory}). We then
define the numerical models (Sec.~\ref{model}), 
and successively analyze dynamic (Sec.~\ref{dynamics})
and 
thermodynamic (Sec.~\ref{statics})
behaviour at thermal 
equilibrium, before exploring 
non-equilibrium glassy states (Sec.~\ref{jamming}).  

\section{Ideal glasses and jammed states}
\label{theory}

In this section we introduce theoretical ideas, tools 
and predictions needed to analyze
the numerical results presented below.

\subsection{Useful analogies for hard and soft particles}
\label{theory-a}
  
It is useful to draw more precise analogies between the glass transition 
observed in thermal glasses and in hard spheres.
In thermal glasses, the natural control parameter
is the temperature, $T$, and observables such a
equilibrium relaxation times, $\tau_\alpha(T)$, and 
the energy density, $e(T)$, are measured. 
In the following, we shall measure relaxation 
times by studying the time decay of the self-part of the 
intermediate scattering function, 
\be
F_s(q,t) = \frac{1}{N} \left\langle \sum_{j=1}^{N} e^{i
{\bf q} \cdot ({\bf r}_j(t)-{\bf r}_j(0))} \right\rangle,
\label{fseq}
\ee
where ${\bf r}_j(t)$
represents the position of particle $j$ at time $t$ in a 
system composed of $N$ particles. The brackets represent an ensemble
average at thermal equilibrium.
In practice we define $F_s(q,\tau_\alpha) = 1/e$ for a fixed wavevector $q$,
corresponding roughly to the first peak of the 
static structure factor and detecting particle motion 
typically over the interparticle distance.
For isotropic pairwise interactions defined by 
the potential $V(r)$, the energy density
reads
\be
e = \frac{1}{N} \left\langle 
\sum_{i=1}^N \sum_{j>i} V(|{\bf r}_{i} - {\bf r}_j|) \right\rangle.
\ee
  
In hard sphere systems,  
the natural control parameter is the volume fraction, 
\be
\phi = \rho \pi \sigma^3 / 6,
\ee
for particles of diameter $\sigma$ and a number density $\rho=N/V$, where
$V$ is the volume of the sample. 
The equation of state of the hard sphere system is then measured
by defining the reduced pressure, 
\be 
Z(\phi) = \frac{P}{\rho k_B T},
\ee
where $k_B$ is the 
Boltzmann constant, and $P$ the pressure. 
In numerical work, the pressure can be measured 
from the Virial, 
\be
Z = 1  -\frac{1}{3 N k_B T} \left\langle 
\sum_{i=1}^N \sum_{j>i} w(  |{\bf r}_{i} - 
{\bf r}_j| ) \right\rangle,
\label{pressMD}
\ee 
where $w(r) = r V'(r)$. 
For the hard sphere system it can conveniently be reexpressed 
as 
\be
Z(\phi) = 1 + 4\phi g(\sigma^+),
\label{pressMC}
\ee
where $g(r=\sigma^+)$ is the pair correlation function
measured at contact.

It is interesting to notice that the volume fraction $\phi$ plays 
a role analogous to the energy density $e$ in molecular glasses, while
the reduced pressure is the  
thermodynamic parameter analogous to temperature.
Since glassiness sets in when $\phi$ and $Z$ increase in 
hard spheres, and when $T$ and $e$ decrease in molecular glasses,
the analogy between both system reads:
\be
T \leftrightarrow \frac{1}{Z}, \quad
e \leftrightarrow \frac{1}{\phi}.
\label{dico}
\ee
Therefore, a finite temperature singularity
in thermal glasses translates into a singularity at 
finite pressure in hard spheres, while the limits $T \to 0$ 
and $Z \to \infty$ are analogous. In both limits, indeed, no 
particle motion is possible.

At the dynamical level, a reference law 
for the relaxation time of molecular glasses is the Arrhenius law, 
\be
\tau_\alpha(T) = \tau_\infty \exp \left( \frac{E}{k_B T} \right),
\ee
where $\tau_\infty$ and $E$ are two constants with dimensions 
of time and energy,
respectively. Deviations from Arrhenius behaviour, `super-Arrhenius
relaxation', are often interpreted as the signature of the 
non-trivial cooperative nature of 
glassy dynamics. This physical intuition can in fact 
be rigorously established using concepts drawn 
from linear response theory~\cite{science,JCP,dalle}. Angell 
has introduced the notion  of `fragility' to quantify these 
deviations~\cite{angell}, 
and suggested to represent the experimental data in an Arrhenius 
representation, $\log \tau_\alpha$ {\it vs.} $1/T$. 
Popular functional forms to account for deviations from Arrhenius
behaviour are the Vogel-Fulcher-Tamman (VFT) law, 
\be
\tau_\alpha(T) = \tau_\infty \exp \left( 
\frac{A}{(T-T_0)^\delta}
 \right), 
\label{vft}
\ee
where the exponent $\delta$ is usually taken to unity, $\delta=1$, 
and the B\"assler law, 
\be
\tau_\alpha(T) = \tau_\infty \exp \left( \frac{B}{T^\alpha} 
\right),
\label{bassler}
\ee
where the value $\alpha = 2$ can be obtained from 
different theoretical perspectives~\cite{bassler,east,pnas,nef}.
An obvious difference between Eqs.~(\ref{vft}) and (\ref{bassler})
is the introduction in the former of a special 
temperature $T_0$ where the relaxation time is predicted to diverge, and
below which a true glass phase should exist.  
In the latter, equilibrium could in principle be maintained
down to $T=0$, and no glass phase exists.  

Using the dictionary in Eq.~(\ref{dico}) we now see that 
these well-known expressions for thermal glasses translate 
into relations between $\tau_\alpha$ and $Z$. 
Thus the simplest dynamic law for hard sphere, analogous
to Arrhenius behaviour, reads 
\be
\tau_\alpha(\phi) = \tau_\infty \exp [ c Z(\phi)], 
\label{arrZ}
\ee 
where $c$ is an adimensional constant. Just as the Arrhenius law
stems from considering relaxation in a liquid arising from a 
thermally activated local relaxation over a fixed energy barrier, 
Eq.~(\ref{arrZ}) describes a relaxation arising from local 
fluctuations of the volume, which justifies fully our analogy between 
$T$ and $1/Z$.
This idea is made more precise 
in free volume approaches, see Sec.~\ref{freevol} below.
 
The VFT and B\"assler expression respectively become
\be
\tau_\alpha(\phi) = \tau_\infty \exp \left(  
\frac{A}{[Z_0-Z(\phi)]^\delta}
\right),
\label{vftZ}
\ee
and 
\be
\tau_\alpha(\phi) = \tau_\infty \exp [B Z^\alpha(\phi)].
\label{basslerZ}
\ee

However, since 
the experimental control parameter is the volume fraction 
$\phi$ while $Z$ is often hard to access experimentally, these
expressions are usually recast in terms of 
relations between $\tau_\alpha$ and $\phi$. This can be confusing 
since the equivalence between soft and hard particles is then 
easily lost, as it depends on the explicit behaviour of the equilibrium
equation of state, $Z=Z(\phi)$. Fitting formula using $Z$ or $\phi$ can only 
take mathematically equivalent forms when
the pressure is finite at the volume fraction where $\tau_\alpha$ 
diverges, if it exists. A well-known example is
the algebraic power laws predicted by 
the mode-coupling theory~\cite{mct} for molecular glasses, 
\be
\tau_\alpha(T) \sim (T-T_{\rm MCT})^{-\gamma},
\label{mctT}
\ee
and for hard spheres, 
\be
\tau_\alpha(\phi) \sim (\phi_{\rm MCT} - \phi)^{-\gamma}.
\label{mctPHI}
\ee
Alternatively, an instance where 
the mathematical equivalence is clearly lost arises if there exists 
a `jamming' density $\phi^\star$
where $Z(\phi)$ diverges, $Z \sim (\phi^\star - \phi)^{-1}$, because 
then an `Arrhenius' behaviour in $Z$ as in Eq.~(\ref{arrZ}) becomes 
similar to a `super-Arrhenius' VFT 
form when the variable $\phi$ is used instead. 
Similarly, strong and fragile molecular glass-forming
materials would look alike if the energy density were used 
instead of the temperature, since both types of system could for instance
behave as $\tau_\alpha \sim \exp(c/(e-e_0))$. 

This example shows that the fragility of hard sphere 
systems should be evaluated by adapting the Angell plot 
to the pressure variable and by plotting $\log \tau_\alpha$ {\it vs.} 
$Z$, emphasizing possible 
deviations from the straight line corresponding to the 
reference law (\ref{arrZ})~\cite{sear}. Note that in a recent 
experimental work, the Angell plot was adapted to soft colloids
using $\phi$ as an abscissa, instead of the pressure~\cite{johan}. 
The above considerations suggest that it would be very 
interesting to reanalyze the behaviour of these 
soft colloids along the lines suggested above in order 
to discuss possible changes of `fragility'.

\subsection{Ideal glass transition}

The above expressions for the relaxation time
in thermal and hard sphere glasses can be justified
from several theoretical approaches. We shall describe those
theoretical frameworks which make specific predictions
for both hard and soft sphere systems, and are therefore 
relevant to the present work.

A finite temperature/pressure dynamic singularity
as in Eqs.~(\ref{vft}) and (\ref{vftZ})
and a true glass phase exist in the context of 
random first order transitions~\cite{KTW,FZ}. 
In the fluid phase  
before the transition, the dynamics is dominated by the
existence of a large number of metastable states, 
allowing the definition of a finite configurational entropy.
The dynamic transition coincides with the point where the configurational 
entropy of the fluid vanishes, and the transition is accompanied 
by a jump in the specific heat (for thermal glasses) or in 
the compressibility (for hard spheres). The glass phase is characterized
by a vanishing configurational entropy, and the structure of phase space is 
that of a system with one-step level of replica symmetry breaking.

The existence of an `ideal' glass transition is an exact
result for a number of many-body interacting models 
defined in mean-field geometries, but remains a
conjecture for three-dimensional realistic particle models. 
Interestingly, it is possible to turn this conjecture into a basis 
for actual microscopic, but approximate, 
calculations of the location of the glass transition
and the structure of the glass phase~\cite{remi,mp}. 
These approximations
have been applied to fluids of hard and soft particles. 
For hard spheres,  
this approach predicts the existence of a divergence of $\tau_\alpha$ at a
finite pressure, $Z_0$, at a packing fraction $\phi_0$~\cite{FZ,silvio,FZ2}. 
Above $\phi_0$ the phase which dominates the 
equilibrium measure is a non-ergodic 
glassy phase characterized by replica symmetry breaking. 
The equilibrium equation of state for the glass phase 
diverges at a packing fraction larger than $\phi_0$, called 
`Glass Close Packing', $\phi_{GCP}$, because it corresponds to the 
densest possible glass with an infinite pressure~\cite{FZ}. 
Replica calculations have not been applied to 
the system of compressible spheres studied below, but they would 
presumably yield a phase diagram as sketched in Fig.~\ref{sketch}-(a), with 
a glass transition line $T_0(\phi > \phi_0)$ emerging from 
point $G$ at $\phi_0$. It would be interesting to 
study also the finite temperature fate of the 
jamming transition at $GCP$ from this 
theoretical perspective. 
 
As mentioned in the introduction, experimental evidence 
relies heavily on extrapolations since equilibrium
can not be achieved very close to the transition. A 
quite favourable experimental finding 
is the coincidence, with reasonable if not decisive 
accuracy~\cite{tanaka}, 
of the extrapolated temperatures 
for the vanishing of the configurational entropy and 
for the divergence of the relaxation time obtained in several
experimental studies 
of molecular glass-formers. For hard particles, this coincidence seems
to hold in numerical studies~\cite{foffi}, 
but the extrapolations are not unambiguous,
and arguably not very convincing.

\subsection{Free volume ideas}
\label{freevol}

Free volume arguments are very popular in the literature of the
molecular glass transition~\cite{freevol}, although they are easier to 
understand in the context of hard spheres. Free volume 
predictions for hard spheres exist both for the equation 
of state and for the dynamical behaviour. In this approach, 
there exists a maximal packing fraction, $\phi^\star$, where
the equilibrium pressure of the hard sphere fluid diverges 
as~\cite{wood}:
\be
Z(\phi) = \frac{d}{1-\phi/\phi^\star},
\label{free}
\ee
where $d$ is the dimension of space. The critical 
packing fraction $\phi^\star$ is often called `random 
close packing' in the literature~\cite{liu}, but we discuss below
its meaning in more detail. 

To obtain dynamical predictions, one then assumes that 
for $\phi < \phi^\star$ each particle
possesses some amount of `free volume', which can then be used to perform 
local relaxations. In the traditional approach, one gets the 
`Arrhenius' prediction in Eq.~(\ref{arrZ}) predicting 
that $\tau_\alpha$ diverges at $\phi^\star$ as 
\be
\tau_\alpha 
\sim \exp(c Z) \sim 
\exp[cd \phi^\star /(\phi^\star - \phi)].
\label{freepred}
\ee 
In this approach, $Z$ and $\tau_\alpha$ diverge together 
at the packing fraction $\phi^\star$, 
in contrast with the ideal glass transition scenario.

Using Eq.~(\ref{dico}), we remark that 
these free volume predictions are then the 
direct analog of the Arrhenius behaviour for thermal glasses, and should
serve as a reference basis to analyze hard sphere data, the analog
of `strong' behaviour for thermal glasses. 
Physically, it indeed makes sense that the divergence in $Z$ 
should be accompanied by a divergence of $\tau_\alpha$ since 
no relaxation is possible when all particles touch each other, just
as no relaxation occurs at $T=0$ for molecular glasses. 
The free volume scenario was very recently revisited 
in Ref.~\cite{ohern}, where $\phi^\star$ was 
given a specific interpretation in terms of the  
jamming transition occuring at point $J$~\cite{pointJ} in the phase diagram
of Fig.~\ref{sketch}-(b), predicting that $\phi^\star = \phi_J$. 
We discuss the physical meaning of $\phi_J$ further below.

In recent work, Schweizer and coworkers have developed a statistical
approach to describe the dynamics of hard spheres at large 
density~\cite{ken}.
They obtain a self-consistent equation of motion for the dynamics 
of a tracer particle which they can solve either numerically, 
or, in some limit, analytically. 
In the latter case, they predict in particular 
a B\"assler law as in Eq.~(\ref{basslerZ}) with $\alpha=2$. 
They assume  
further that the pressure diverges at random close packing
as in Eq.~(\ref{free}), producing therefore the prediction 
of a non-trivial `fragile' behaviour for hard spheres, namely
\be
\tau_\alpha \sim \exp( B Z^2) \sim \exp[Bd (\phi^\star)^2
/(\phi^\star - \phi)^2].
\label{KS}
\ee

Applying this body of ideas to a system of compressible spheres, 
we obtain the phase diagram sketched in Fig.~\ref{sketch}-(b). Here,
the hard sphere system is ergodic up to the maximal 
volume fraction where the equilibrium pressure diverges and the system jams, 
and no ideal glass transition occurs. At finite temperature, dynamic arrest
only occurs in the limit where $T \to 0$ and a true glass phase 
never exists at finite temperature. This phase diagram 
is consistent with the physical idea that
no glass transition 
can occur for thermal glasses at any finite temperature, because 
it is always possible to perform `local' relaxations at a finite 
energy cost. Equivalently for hard spheres, 
one states that as long as some amount 
of free volume is available to each particle, it is possible to 
relax by appropriately displacing a finite set of particles. 
This simple argument, which can be made more formal~\cite{osada}, 
provides evidence that 
the self-diffusion constant for hard spheres does not vanish
at finite pressure, 
but it says nothing about collective relaxation timescales, which 
can still diverge. Using (with no justification) 
this argument against collective 
freezing, one would conclude that 
the glass phase only exists along the $T=0$ line,
as in Fig.~\ref{sketch}-(b). Note that
self and collective relaxations often 
yield similar results,
suggesting that the formal argument in 
Ref.~\cite{osada} is not very useful for accessible 
timescales.

One could easily imagine scenarios intermediate between 
the two sketches in Fig.~\ref{sketch}. 
For instance, it is sometimes stated that no glass transition 
can occur for thermal glasses at any finite temperature, 
but that hard spheres are somewhat different because of 
the hard constraint imposed by the potential~\cite{proca}.
In that case, a glass phase would never exist at finite temperature, 
but the glass line at $T=0$ could extend down to a low-density limit 
occuring at a finite pressure, so that $G \neq J$. 
Note however that since the proof presented in 
Ref.~\cite{proca} applies to finite
size systems, $N < \infty$, it does not directly contradict the existence
of an equilibrium phase transition in the thermodynamic limit $N \to \infty$,
as sketched in Fig.~\ref{sketch}-(a). 
 
The final alternative is of course the absence 
of a glass of hard spheres before jamming along the $T=0$ line,
together with a glass transition line emerging from random 
close packing $T_0(\phi \to \phi^\star)=0$. 
This last scenario would actually be consistent
with free volume models developed for both hard and soft particles 
which does predict a VFT divergence at finite temperature 
for molecular glass-formers where the amount of `free volume'
disappears~\cite{grest}, 
but no such glass transition in the case of hard spheres
where free volume is available up to infinite pressure.

\subsection{Jamming transitions and random close packing} 
\label{rcp}

The glass transition described above locates
state points where ergodicity is lost. For assemblies of 
spherical particles, another transition occurs at large density
which can be defined in purely geometric terms, without
invoking thermal equilibrium or ergodicity~\cite{pointJ,torquato}. 
This jamming
transition is closely related to the notion of
random close packing, very widely discussed experimentally in the 
context of granular materials~\cite{rmpgrains}. 

Granular experiments provide the simplest description of 
a jamming transition for hard particles. Due to gravity, 
a disordered assembly of grains will settle in the bottom of a container
until contact constraints prevent any further 
displacements. The simplest question that can be asked is: what
is the volume fraction occupied by the grains in this jammed state?
Turning to frictionless spheres without gravity, 
the problem is therefore to produce assemblies of particles  
that can not be compressed further without allowing overlaps 
between particles~\cite{pointJ,torquato2,roux,donev}. 
These states are therefore 
infinite pressure hard sphere configurations. Note that
the reverse is not true, because not all infinite pressure state 
are jammed~\cite{donev}. 

Since several models for the dynamics of hard spheres predict
that $\tau_\alpha$ and $Z$ diverge together at the same 
density, it is very natural to conjecture that glass and jamming
transitions may be two facets of the same transition~\cite{liunagel}, 
which could therefore be explained in purely geometric 
terms~\cite{matthieu}. In that case, random close packing would correspond
to the diverging point in the equilibrium
equation of state for hard spheres, and would in effect control 
the glassy dynamics of colloidal particles. 

The situation gets more complicated if an ideal glass transition 
occurs~\cite{FZ,jorge1,jorge2} 
because the equilibrium pressure does not diverge at the density
where ergodicity is lost, and so jamming and glass transitions
are decoupled, as sketched in Fig.~\ref{sketch}-(a). In that case, 
the energy/volume fraction analogy in Eq.~(\ref{dico}) 
suggests that finding the 
glass close packing density is 
equivalent to finding the energy of the ground state in 
a disordered system with a complex energy landscape~\cite{jorge1}. 
This is likely
a computionally hard problem. This suggests that any `physical' 
algorithm used to produce $T=0$ or $Z=\infty$ states
will end up in configurations that have an energy larger than 
the true ground state (in thermal glasses), or that will jam at a density 
smaller than the glass close packed state (in hard spheres), because
they will remain trapped within the metastable states that 
are responsible for the glassy behaviour at low temperature/large
density.

A final complication is the 
existence for both hard and soft particles of a first 
order transition towards a crystalline structure, 
which is ignored in the above-mentioned approaches.
The existence of the crystal phase might render the fluid metastable, 
and crystallization can indeed occur during experiments or simulations
for systems with low glass-forming ability. It is however easy to 
bypass crystallization in experiments for a large number of molecular
systems, and for hard spheres using a sufficient amount of size 
polydispersity. 
If crystallization does not occur (which can be checked by monitoring the 
structure of the fluid), it is then possible to study the metastable 
fluid phase at `thermal equilibrium', and it is the situation usually
considered by theoretical approaches to the glass transition.
In that case, one applies concepts from equilibrium 
thermodynamics to study the system in a reduced 
part of his phase space corresponding to metastable 
disordered states, the crystal region being excluded.

In the same vein, if jamming is studied 
by driving hard sphere configurations out of equilibrium using 
specific `physical' algorithms for compressions, 
then again crystallization is not a critical issue for sufficient
polydispersity. However, it should
be kept in mind that a large number of crystalline and polycrystalline
configurations can be built over a broad range of densities encompassing 
the putative glass phase, 
for instance by artificially mixing 
crystalline and fluid phases~\cite{mixture}. 
The existence of the crystal thus makes the 
definition of a unique ground state for thermal glasses, or the maximum 
packing fraction for hard spheres `ill-defined'~\cite{torquato}, 
because these concepts are not made in reference to thermal equilibrium.
Below, we shall study jamming transitions 
specifically designing algorithms for which
crystallization and demixing are under controll.

To make this argument useful to analyze the (im)possibility of 
an equilibrium glass transition~\cite{mixture}, 
one should additionally show that nucleation of these 
ordered states from the disordered metastable fluid is indeed possible 
at thermal equilibrium. Even then, if nucleation barriers happen to be 
extremely large, then the glass transition might indeed be cutoff 
at extremely large relaxation timescales by nucleating the crystal
phase rather than undergoing an ideal glass transition, but the argument
is thus not necessarily useful for experimentally 
accessible timescales.  

\section{Models for hard and soft particles}
\label{model}

We have explored the equilibrium behaviour and glassy states 
in two numerical models for hard and soft repulsive spheres, which we 
now introduce.

\subsection{Hard spheres}

We use a standard Monte Carlo algorithm~\cite{allen} to study
numerically a 50:50 binary mixture of hard spheres with  diameters
$\sigma$ and $1.4 \sigma$, known to efficiently prevent
crystallization~\cite{pointJ}. 
We work in a three dimensional space with periodic
boundary conditions, and mainly use $N=1000$ particles when 
studying thermal equilibrium. No
noticeable finite size effects were found in runs with
$N=8000$ particles performed for selected state points. 
We have detected no sign of crystallization or demixing between 
large and small particles in all our simulations, some of 
which having run for more than $10^{10}$ Monte Carlo steps.  

In an elementary move, a particle is chosen at random and assigned a
random displacement drawn within a cubic box of linear size
$0.1\sigma$ centered around the origin. The move is accepted if the
hard sphere constraint remains satisfied. One Monte Carlo step
corresponds to $N$ such attempts. 
Comparison of Monte Carlo dynamics with more standard 
Molecular Dynamics simulations of glass-forming liquids have 
provided evidence that slow relaxation in dense fluids is insensitive 
to the choice of a microscopic dynamics~\cite{MC1,MC2}. 
This is by no means a trivial result, as both types of dynamics 
could in principle yield widely different results for the dynamic
behaviour, especially 
in those cases where collective particle 
motions are believed to play an important role.

We have also performed non-equilibrium compressions 
to explore glassy states at large volume fraction. In that case
we have systematically used two system sizes, 
$N=1000$ and $N=8000$, and the reported data for compressions 
are those for the larger system, although they are statistically 
indistinguishable from those obtained for the smaller system. 
To perform a compression, we use the following procedure. We start from 
an equilibrated hard sphere fluid configuration at a given $\phi$. We then 
perform an instantaneous compression of the simulation box, which produces
overlaps between particles, which are removed using the above Monte Carlo 
algorithm. As soon as all overlaps have disappeared, we 
perform the next compression of the system. We adjust the compression 
rate to maintain the number of overlaps after each compression below 
a constant 
number, $0.025N$. When density gets large, it becomes difficult to remove all 
overlaps and we stop the compression when at least one overlap 
has survived after $10^5$ Monte Carlo steps. 
For these non-equilibrium compressions, we have no insurance 
that Monte Carlo and Molecular Dynamics yield comparable results. 
This is not crucial for our purposes because we do not wish to perform 
dynamic measurements at large volume fraction. 

\subsection{Harmonic spheres}

We use Molecular Dynamics simulations~\cite{allen} to 
study a system composed of compressible 
particles~\cite{durian,pointJ}
interacting through a pair-wise potential: 
$V(r_{ij}) = \epsilon (1- r_{ij}/\sigma_{ij})^2$ for
$r_{ij} < \sigma_{ij}$, $V(r_{ij})=0$ 
otherwise. The interparticle distance is $r_{ij} = |{\bf r}_i
-{\bf r}_j|$ and $\sigma_{ij} = (\sigma_i + \sigma_j)/2$,
where ${\bf r}_i$ and $\sigma_i$ are the position and diameter 
of particle $i$, respectively. 
We shall use the term `harmonic spheres' 
for this system because the repulsive
force between spheres is linear in their overlap. 
We use system sizes between 
500 and 8000 particles, and report 
the results obtained for $N=1000$, for which no finite size effects 
are detected, within numerical accuracy.

We prevent crystallisation by using the same
50:50 binary mixture of spheres of diameter ratio 
$1.4$ as used in the hard sphere case.
Up to volume fraction $\varphi = 0.846$ 
we detect no sign of crystallization at all
studied temperatures; at and above $\phi = 0.924$ there was 
evidence of incipient crystallization at the lowest 
temperatures~\cite{tom}.
However, these crystallization effects occur well away from 
the region of interest around point $G$ in the regime 
$\phi = 0.55 - 0.75$.

\begin{figure}
\psfig{file=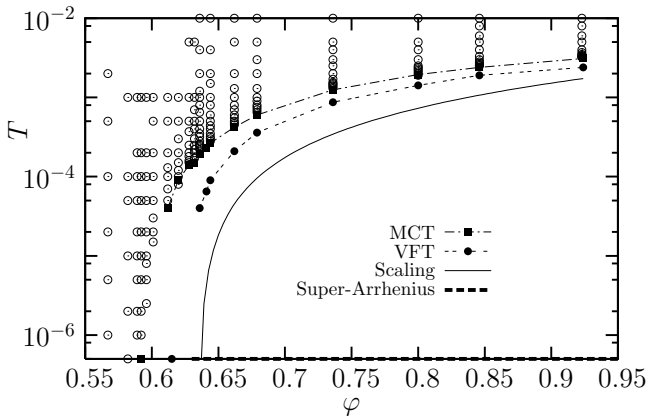,width=8.5cm}
\caption{\label{phaselong} Phase diagram for harmonic spheres. 
Open symbols represent the investigated state points. 
The location of glass lines obtained by 
fitting to known functional form (MCT, VFT), and using 
activated scaling (\ref{glassline}) are shown, while the  
`super-arrrhenius' line corresponds to volume fractions 
where $\tau_\alpha(\phi,T)$ increases faster than an 
Arrhenius law when $T$ is reduced at constant $\phi$.}
\end{figure}

We use $\epsilon$ as the energy unit, and 
$\sqrt{\sigma_2^2/\epsilon}$ as time unit, with masses set to unity.
All dynamical 
results are obtained at thermal equilibrium, which has been carefully 
controlled. When temperature is low and density is large, we are not
able to thermalize.
Crystallization and equilibrium issues determine the 
boundaries of the 
region of investigated state points shown 
in Fig.~\ref{phaselong}.

Finally we have performed exploration of glassy states of hard spheres
using the compressible sphere system as follows. We start from 
an equilibrated configuration of harmonic spheres at a given state point
$(\phi,T)$. We then rapidly cool the system at constant density down to 
$T \to 0$ using Molecular Dynamics. We track pressure and energy during 
compression and obtain valid hard sphere configurations at $T=0$ 
when energy vanishes and $Z$ remains finite and independent of $T$ 
at sufficiently low temperatures. It is equal to the value of the pressure 
for the corresponding hard sphere system.

\section{Evidence for a dynamic singularity for hard spheres}
\label{dynamics}

In Refs.~\cite{luca,lucalong} we analyzed in detail the 
volume fraction dependence of the relaxation time for hard spheres, 
while Ref.~\cite{tom} contains a discussion of the interplay between
density and temperature for harmonic spheres. Thus, in this section, 
we summarize the main conclusions drawn in these papers, and
complement them by further analysis of these dynamic measurements. 
 
\subsection{Fitting to known functional forms}

For both hard and harmonic spheres, we find that 
algebraic divergences as predicted by mode-coupling theory
only hold over a restricted time window of about 2 to 3 decades
in the regime comprising the onset of glassy dynamics. This is 
consistent with findings in other 
systems~\cite{mct}.  
First, fitting the relaxation time $\tau_\alpha(\phi)$ for hard spheres
to Eq.~(\ref{mctPHI}) yields a mode-coupling 
singularity at $\phi_{\rm MCT} \approx 0.592$
with a critical exponent $\gamma \approx 2.6$, as reported in 
Ref.~\cite{luca}. This is consistent with previous analysis of hard
sphere systems~\cite{szamel,puertas}.

Second, for harmonic spheres, we fitted the temperature 
evolution of $\tau_\alpha(\phi,T)$ at constant $\phi$
between $\phi=0.61$ and $\phi=0.924$ to Eq.~(\ref{mctT}). 
We thus obtain a mode-coupling transition line, 
$T_{\rm MCT}(\phi)$, as shown in Fig.~\ref{phaselong}. 
Our data are consistent with a fitted MCT temperature which vanishes
rapidly as $\phi$ decreases towards $\phi_{\rm MCT}$, although 
it is difficult to obtain an accurate determination of $T_{\rm MCT}$
very close to $\phi_{\rm MCT}$. We find also that the critical exponent 
$\gamma$ in Eq.~(\ref{mctT}) has a strong volume fraction 
dependence, increasing from $\gamma \approx 2.8$ for $\phi=0.612$ to  
a value $\gamma \approx 5.3$ for $\phi=0.8$ and above.   
Although the algebraic divergence predicted by MCT is eventually 
avoided, as we shall describe shortly, it would be very interesting 
to analyze the dynamic behaviour of the harmonic sphere system 
using mode-coupling theory: does theory reproduce the strong density
dependence of the critical exponent, and does it predict 
specific scaling properties in the vicinity of the
hard sphere point at $T=0$ and $\phi=\phi_{\rm MCT}$?  

Since deviations from the algebraic divergence predicted by MCT 
are observed  at low enough temperatures, we repeated our data
analysis using standard empirical 
approaches. First, we have fitted our data for hard spheres
using a VFT form,
\be 
\tau_\alpha(\phi) \sim
\exp \left[ \frac{A}{(\phi_0-\phi)^\delta} \right], 
\label{activatedHS}
\ee
by analogy with Eq.~(\ref{vft}). 
Imposing the standard value $\delta=1$, one locates a
critical volume fraction at $\phi_{\rm VFT} \equiv 
\phi_0(\delta=1) \approx 0.615$. However,
using $\delta$ as an additional fitting parameter, 
a slightly better fit is obtained for $\delta \approx 
2.2$ from which a larger critical packing fraction, 
$\phi_0(\delta=2.2) \approx 0.635$, is estimated. Remarkably, similar
conclusions hold for experiments performed on colloidal hard 
spheres~\cite{luca,lucalong}.

We have then fitted our finite temperature data for harmonic spheres 
to the VFT form in Eq.~(\ref{vft}), imposing an exponent 
$\delta=1$ at all densities. We find that such a fit can describe our 
data at all volume fractions rather well for $\phi>0.636$, and report 
the VFT glass line deduced from this fitting procedure in 
Fig.~\ref{phaselong}. 
Again it becomes difficult to obtain accurate determination of the VFT 
temperature as $\phi$ decreases, but the extrapolation
of the VFT line is nevertheless in good agreement with the hard sphere
result $\phi_{\rm VFT} \approx 0.615$.

An interesting outcome of the VFT fitting procedure is the 
density dependence obtained for the fitting parameter $A$ 
in Eq.~(\ref{vft}). For $\delta=1$, it is convenient
to define the parameter $D \equiv A T_0$, which serves as
an experimental tool to quantify the fragility of supercooled 
liquids~\cite{angell}. 
We find that $D$ varies strongly with $\phi$ and changes 
from $D \approx  55$ for $\phi=0.636$ down to 
$D \approx 8$ for $\phi>0.80$. In experimental investigations 
of the dynamics of supercooled liquids~\cite{angell,tanaka}, 
$D$ is found to decrease
similarly from $D\approx 60$ for SiO$_2$, a strong glass-forming
material, down to $D\approx 30$ for liquids of intermediate 
fragilities such as glycerol or ZnCl$_2$, and to $D\approx 10$ for 
fragile liquids such as orthoterphenyl (the fit to a VFT form
is obviously more ambiguous for strong glass-forming materials with 
nearly Arrhenius behaviour). Thus, we find that the system 
of harmonic spheres displays a variation in kinetic 
fragility which encompasses the range observed in experiments, 
a central claim made in Ref.~\cite{tom}, which is confirmed by the 
present analysis.   

Both MCT and VFT fitting formula contain a divergence 
at a critical temperature, corresponding to the phase diagram sketched 
in Fig.~\ref{sketch}-(a). However,  
to explore the possibility of a $T=0$ glass line at large 
density, as sketched in Fig.~\ref{sketch}-(b),  
we have also fitted our data to a generalized B\"assler form,
as in Eq.~(\ref{bassler}), using $B$ and $\alpha$ as free 
fitting parameters for each volume fraction. We find that 
Eq.~(\ref{bassler}) also describes our data rather well, but we must 
use an exponent $\alpha$ which 
increases from 1 near $\phi \approx 0.63$ up to $\alpha \approx 
3.7$ at $\phi=0.736$, while $\alpha=2$ is 
traditionnally preferred in supercooled liquids~\cite{bassler,fit3}.
The range of density where $\alpha > 1$ 
delimits the density region where `super-Arrhenius' behaviour
is observed, and is indicated with a dashed line in Fig.~\ref{phaselong}.

Therefore, we conclude cautiously that fitting
our data to known functional forms shows that above 
$\phi \approx 0.63$, the dynamics at constant $\phi$ slows 
down with $T$ faster than an Arrhenius law, and exhibits 
dynamics typical of fragile glass-forming liquids, with a fragility
increasing dramatically with $\phi$.
This fragility increase is accompanied by a large variation
of the MCT critical exponent $\gamma$, of the VFT fitting 
parameter $D$, and the exponent $\alpha$ in Eq.~(\ref{bassler}), 
but these fitting procedures leave open the location of the 
divergence of $\tau_\alpha$, since both finite temperature 
and singularity-free fitting formula can be used to describe our data.
Similarly, the activated form in Eq.~(\ref{activatedHS}) 
is clearly favoured by our hard sphere data, 
possibly with a non-trivial exponent $\delta >1$, but 
the location of $\phi_0$ has to be extrapolated from the analysis 
of much smaller volume fractions, and must be discussed with caution.

\subsection{Activated scaling near point $G$}
\label{scaling}

The weakness of the above analysis was mentioned 
in the introductory lines of this article. When 
independently fitting a single data set obtained by changing a single control 
parameter, the range of timescales
covered  by simulations (and experiments!) is 
usually too small to discrimate
between very different fitting formula.

In Ref.~\cite{tom} we suggested to apply ideas from 
dynamic scaling to the data obtained in the whole 
$(\phi,T)$ plane of harmonic spheres to gather more precise 
information on the phase diagram, and in particular the location of 
the dynamic singularity at point $G$. To the best of our knowledge, 
such an analysis using two control parameters has no counterpart in the 
glass transition literature. 

Our main aim is to determine the location $\phi_0$ 
of point $G$ along the 
$T=0$ hard sphere axis starting from the following qualitative
considerations about the harmonic sphere system. 
For $\phi < \phi_0$, the dynamics slows down 
when $T$ decreases, but the relaxation time saturates in the limit 
$T \to 0$ to a finite value corresponding to the hard sphere fluid.  
For $\phi > \phi_0$, however, the relaxation time should 
extrapolate to infinity in this limit, by definition of $\phi_0$. 
These two regimes are obviously delimited by $\phi=\phi_0$, 
where the system, like Buridan's ass,
`hesitates' forever between these two regimes.
These three different situations are all included 
in the following scaling form~\cite{tom}:
\be
\tau_\alpha(\phi,T) \sim
\exp \left[ 
\frac{A}{|\phi_0 - \phi|^\delta} F_{\pm} \left(
\frac{ |\phi_0 - \phi|^{2/\mu} }{T} 
\right)
\right] .
\label{critical}
\ee
In this expression, $F_{\pm}(x)$ are scaling functions
applying to volume fractions above and below $\phi_0$, respectively.
We expect therefore that $F_-(x \to \infty) \to 1$ to recover the
hard sphere fluid limit, Eq.~(\ref{activatedHS}), when 
$T \to 0$ and  $\phi < \phi_0$. 
Similarly, $F_+(x \to \infty) \to \infty$, for
$\phi > \phi_0$. Moreover, 
continuity of $\tau_\alpha$ at finite $T$ when crossing
$\phi=\phi_0$ implies a common limit for both scaling 
functions in the $x \to 0$ limit: $F_-(x\to 0) \sim F_+(x \to 0) 
\sim x^{\delta \mu / 2}$, so that 
$\tau_\alpha(\phi=\phi_0,T) \sim \exp(A / T^{\delta \mu /2})$. 

Dynamic scaling was recently observed for athermal jamming 
transitions~\cite{teitel,hatano}, but the nature of the 
hard sphere divergence (algebraic instead of exponential)
was qualitatively different from Eq.~(\ref{critical}), and the 
critical density appearing in the scaling formula also had a different
nature, since these data were not collected at thermal equilibrium.

\begin{figure}
\psfig{file=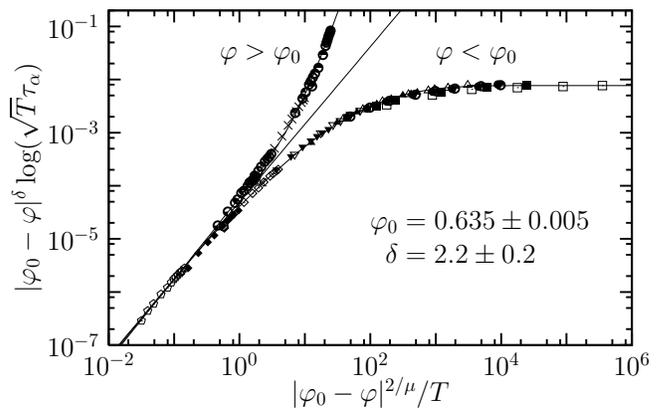,width=8.5cm}
\caption{\label{epl} 
Activated dynamic scaling of relaxation timescales for 
$\phi \in [0.567, 0.736]$. The data for $\phi<\phi_0$ and $\phi>\phi_0$ 
collapse on two distinct branches,  
as described by Eq.~(\ref{critical}).
Times are rescaled by $1/\sqrt{T}$ so that 
the $T \to 0$ limit coincides with hard spheres thermalized at $T=1$.
The errorbars describe the range of values for which 
acceptable data collapse is obtained; 
$\mu=1.3$ is fixed using potential energy considerations.}
\end{figure}

Using Eq.~(\ref{critical}), data at all temperatures $T$, 
and for volume fractions in the range $\phi \in [0.567, 0.736]$
can be collapsed onto the two expected 
scaling branches and the best data collapse is obtained 
for $\phi_0 = 0.635$, $\delta=2.2$ and $\mu=1.3$. 
This is shown in Fig.~\ref{epl}, which reproduces 
the data collapse presented in Ref.~\cite{tom}.
Remarkably, 
while the exponent $\delta$ is not very much constrained by the 
hard sphere data alone, a value below $\delta \approx 2$ 
cannot be used if data 
collapse is sought for harmonic spheres at finite temperatures.
Fixing $\delta$ near 2 thus 
allows us to estimate $\phi_0$ is a much more precise manner:
\be
\phi_0 = 0.635 \pm 0.005,
\label{pointG}
\ee
where the errorbars refer to the range of volume fractions
for which acceptable data collapse is obtained.

The value of the exponent $\mu$ and the form of the scaling variable 
in Eq.~(\ref{critical}) were discussed in terms of an 
effective hard sphere radius in Ref.~\cite{tom},
and the consequences
on the strong volume fraction dependence of the glass fragility
explored in some detail.

While the location of point $G$ is well established 
by the scaling analysis suggested by Eq.~(\ref{critical}), 
the nature of the phase diagram for $\phi > \phi_0$ is 
not completely determined since it depends on the specific form 
of the scaling function $F_+(x)$ for large values of its argument. 
If $F_+(x)$ diverges at a finite value $x_0$, say
$F_+(x) = x^{\mu\delta/2} /(x_0^{\mu/2}-x^{\mu/2})^\delta$, then
Eq.~(\ref{critical}) would yield a glass line of the form 
\be
T_0(\phi) \sim (\phi-\phi_0)^{2/\mu},  
\label{glassline}
\ee
which is shown as a full line in Fig.~\ref{phaselong}. 
However, we checked that our data along the $F_+$ branch can also be
described using a non-diverging form, 
$F_+(x) = x^{\mu\delta/2}(1+b x^{\beta})$, which would be consistent
with a $T=0$ glass line above $\phi_0$ together with the 
B\"assler form in Eq.~(\ref{bassler}) with a density dependent 
effective exponent 
$\alpha$.
Therefore, we conclude that our scaling analysis
leaves open the existence of a finite temperature singularity above
$\phi_0$. 

We emphasize that our conclusion that the equilibrium
relaxation times for hard spheres 
diverge at the volume fraction (\ref{pointG}) 
relies on a demanding scaling analysis of a large set of data in the 
$(\phi,T)$ phase diagram of harmonic spheres, combined with the 
analysis of the $T=0$ hard sphere axis over seven decades
of relaxation times. 
However, as is unavoidable in this field, 
we should not exclude that a different dynamic regime can be entered 
when relaxation timescales beyond reach of our numerical
capabilities are added to the analysis, thereby 
asymptotically changing the overall 
picture presented in this work. If so, however, 
this putative new regime would be experimentally irrelevant for 
colloidal particles~\cite{luca,lucalong}.
  
\section{Equilibrium pressure data}
\label{statics}

The above results support the existence of a non-trivial divergence of
the relaxation time for hard spheres at a critical volume 
fraction $\varphi_0$. 
Quite generally, a divergence is
expected if the equilibrium reduced pressure $Z (\varphi)$
also diverges, because no particle motion is 
possible in this limit, which is analogous to $T=0$ for
systems with soft potentials, cf. Eq.~(\ref{dico}). 
In this section, we ask whether
$\phi_0$ coincides with a divergence in the presssure, and whether
relaxation times can be related to pressure in a direct manner.

\subsection{Results for hard spheres}

\begin{figure}
\psfig{file=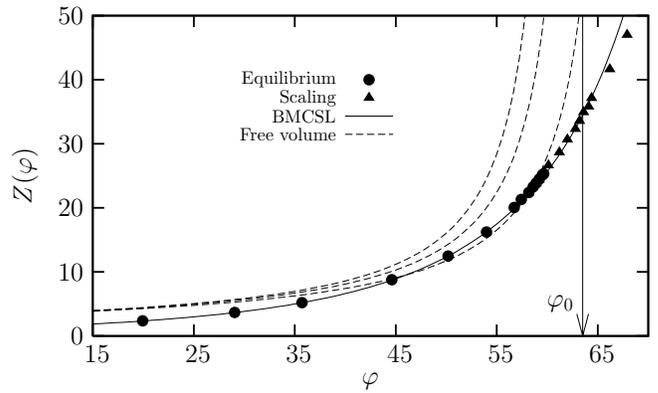,width=8.5cm}
 \caption{\label{presseq} Equilibrium pressure for hard spheres obtained
from direct Monte Carlo simulations (circles),  
and extended to large $\phi$ using the scaling 
behaviour of the harmonic sphere system 
(triangles). The free volume prediction (\ref{free}) is presented
for three locations of the divergence: at $\phi_{\rm VFT} =0.615$, 
$\phi_0=0.635$, and the `best' fit $\phi^\star = 0.672$, and none of them
accurately describes the data. Instead, the BMCSL equation of state 
from liquid state theory describes
the data very well over the entire fluid range and leave the 
pressure finite at $\phi_0$, where the relaxation time diverges.}
\end{figure}

We first describe the equilibrium data obtained from 
direct equilibrium simulations of hard spheres, from which 
pressure is measured through Eq.~(\ref{pressMC}). The results 
are shown as filled circles in Fig.~\ref{presseq} from
very low volume fractions where $Z \approx 1$ up to the largest volume 
fraction for which equilibrium could be reached, $\phi=0.597$, where 
$Z \approx 25.3$.
 
As recalled in Sec.~\ref{freevol}, free volume arguments 
predict a simple form for the divergence of pressure as $Z \sim 
(\phi^{\star} - \phi)^{-1}$.  
In Fig.~\ref{presseq} we attempt a description of our equilibrium 
pressure data for three values of $\phi^\star$. We impose 
$\phi^{\star} = \phi_{\rm VFT} = 0.615$, the volume fraction deduced
from a VFT fit to the dynamic data, and $\phi^\star = \phi_0 = 0.635$, 
our best estimate for the location of point $G$. These fits are clearly
inconsistent with the data, as they do not even go through any of the data
points. This directly implies that the free volume prediction 
in Eq.~(\ref{freepred}) incorrectly represents our data. 
In the same vein, our pressure data are inconsistent 
with a free volume divergence of the equilibrium pressure 
at $\phi_0$, and we do not know how to extrapolate 
$Z(\phi)$ to obtain a diverging pressure at point $G$. 
Thus we conclude that point $G$ defined from the study of the
equilibrium dynamics, does not seem to correspond to point $J$
defined from a pressure divergence, in contrast with a recent 
proposal based on a percolation approach~\cite{ohern}.

If we insist that Eq.~(\ref{free}) must describe at least the last
data points obtained at large volume fraction we find that the value 
$\phi^\star \approx 0.672$ represents the `best' compromise, as 
shown in Fig.~\ref{free}. Clearly, however, the shape of the pressure is not 
very well reproduced. Therefore, our results are in disagreement with 
those obtained in Ref.~\cite{liu}, where the equilibrium 
equation of state 
for hard spheres was fitted using a free volume expression. We believe that
the discrepancy stems from the fact that non-equilibrium 
pressure data obtained in fast compressions were 
incorrectly mixed with equilibrium data and included into the 
free volume fit.

Finally, we show in Fig.~\ref{presseq} that the so-called 
Boublik, Mansoori, Carnahan, Starling and Leland
equation of state (BMCSL)~\cite{bmcsl1,bmcsl2}, 
which is the extension 
to binary mixtures of the Carnahan-Starling equation of state 
for monodisperse 
hard spheres, describes our equilibrium data very accurately over the 
entire fluid range up to $\phi=0.597$. 
A similarly good agreement is known to occur when using the 
Carnahan-Starling equation of state applied to 
monodisperse systems~\cite{hansen}, up to large volume fractions. 
This excellent agreement is a remarkable result, since the BMCSL 
is a reasonable, but somewhat empirically derived, 
equation of state obtained from integral equations
using uncontrolled approximations.
An important feature of the BMCSL equation of 
state, in the present context, is that it is a continuous and derivable 
function up to very large volume fraction, and it 
predicts a singularity in the 
pressure at an unphysically large volume fraction, $\phi=1$.
Therefore, the BMCSL equation of state predicts neither 
a glass transition nor a jamming transition, and we cannot 
use it to extrapolate critical values of the volume fraction
for the hard sphere system.

\begin{figure}
\psfig{file=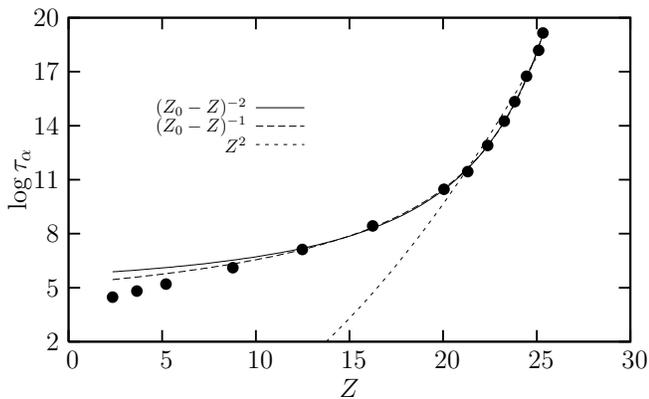,width=8.5cm}
\caption{\label{angell} `Angell plot' for hard spheres showing the pressure 
dependence of $\log \tau_\alpha$ where the `Arrhenius' behaviour predicted 
by free volume arguments should appear as a straight line. Hard spheres 
instead 
display `fragile' behaviour which is not well fitted either 
by a B\"assler law,
Eq.~(\ref{basslerZ}) with $\alpha=2$,  
but is consistent with a finite pressure singularity. Excellent fits to 
Eq.~(\ref{vftZ}) with $(\delta=1, Z_0=30.2)$
and $(\delta=2, Z_0=34.4)$ are shown.} 
\end{figure}
  
In Fig.~\ref{angell} we present an `Angell plot for hard spheres',
that is, we show the pressure evolution of the logarithm
of the relaxation time. In this plot, the Arrhenius behaviour 
Eq.~(\ref{arrZ}) should appear as a straight line, suggesting 
a simple mechanism for the glassy dynamics of hard spheres.    
Clearly, the data in Fig.~\ref{angell} do not follow such a simple law, 
and hard spheres thus behave in a non-trivial manner: they are 
`fragile', in the precise sense defined in Sec.~\ref{theory-a}. 
As discussed above, this is in contrast with free volume~\cite{freevol}
and percolation-based~\cite{ohern} predictions 
for the dynamics of hard spheres. 

We must thus turn to `fragile' predictions for the behaviour of 
hard spheres as a function of pressure. We first test 
the prediction by Schweizer and coworkers~\cite{ken} 
in Eq.~(\ref{KS}). We find that
a fit of $\log \tau_\alpha$ with $Z^2$ is rather poor. 
In fact, a plot of $\log \tau_\alpha$ vs. $Z^2$ does not linearize the data,
so the quadratic fit in Fig.~\ref{angell} is somewhat arbitrary. 
Moreover, the fit shown in Fig.~\ref{angell}, obtained by focusing 
on the data at larger $\phi$ yield a 
microscopic attempt time 
$\tau_\infty \sim 8\cdot 10^{-3}$ (in MC step units), which is 
physically much too small since it is more than 4 orders of magnitude 
smaller than the relaxation time obtained in the low density limit, 
$\tau_\alpha (\phi \to 0) \approx 10^2$. 
We conclude therefore that the 
(asymptotic) expression in Eq.~(\ref{KS}), although 
predicting the correct upward curvature in the Angell plot in 
Fig.~\ref{angell}, is not an accurate representation of our data.  
Thus, the near coincidence between the exponent $\delta \approx 2.2$
in Eq.~(\ref{activatedHS}) and the B\"assler expression in Eq.~(\ref{KS})
is fortuitous. In fact, using Eq.~(\ref{basslerZ}) leaving $\alpha$ 
free to take values different from 2, we find that the 
data are best described, at large density, using $\alpha \approx 6$, 
suggestive of a pressure dependence of the data which is 
much stronger than the one predicted by Eq.~(\ref{KS}). 

Therefore, we explore a final possibility, suggested by the pressure 
data in Fig.~\ref{presseq}, of an equilibrium pressure which 
actually stays finite when the relaxation time diverges, 
as in Eq.~(\ref{vftZ}). As can be seen in Fig.~\ref{angell},
our data are indeed well described by a finite
pressure singularity, although the hard sphere
data themselves are equally well fitted using $\delta=1$ or $\delta=2$, 
as shown in Fig.~\ref{angell}. This is expected since 
the volume fraction dependence of $\tau_\alpha$ for hard spheres 
was also well fitted by Eq.~(\ref{activatedHS}) 
with $\delta=1$ and $\delta=2$, the latter yielding a 
marginally better fit. To conclude with the pressure, we recall that
a value close to $\delta=2$ was favoured through the analysis 
of the dynamic data for harmonic spheres.  Therefore, our best 
estimate for the critical 
pressure $Z_0$ of the hard sphere system is obtained 
from the fit of the pressure with $\delta=2$ 
shown in Fig.~\ref{angell} in the range
$Z>10$ and yields:
\be
Z_0 \approx 34.4 \pm 0.4,
\label{Z0}
\ee
with errorbars as given by the fitting numerical routine. 
 
\subsection{Pressure results for harmonic spheres}

Just as investigating the dynamics of harmonic spheres in the 
vicinity of point $G$ allowed for an accurate determination 
of the critical density $\phi_0$ 
for the divergence of the relaxation time for hard
spheres, we can use the harmonic spheres to confirm
the above finding that the equilibrium pressure of hard sphere 
is finite at $\phi_0$, and obtain an independent determination 
of $Z_0$, which does not rely on the value of 
other fitting parameters. 

\begin{figure}
\psfig{file=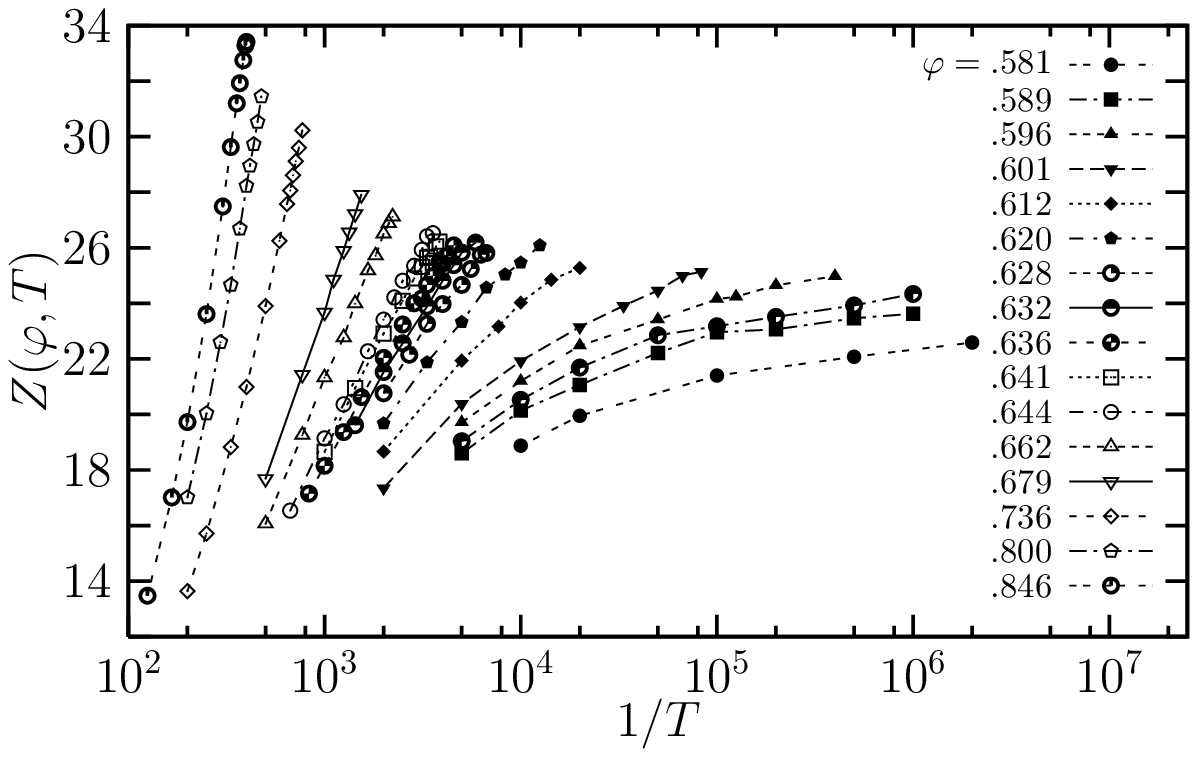,width=8.5cm}
\psfig{file=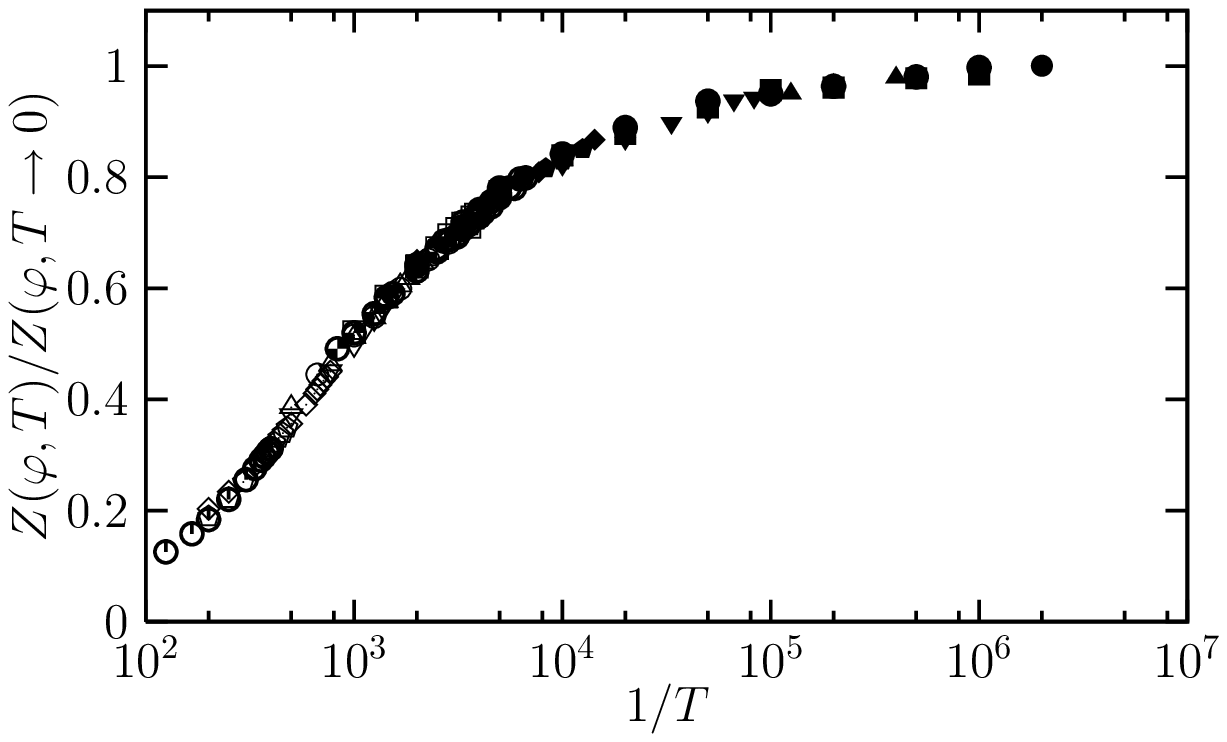,width=8.5cm}
\caption{\label{pressure} Top:
Temperature dependence of the pressure of harmonic spheres 
at different volume fractions. 
Bottom: rescaling the pressure by the factor $Z(\phi,T\to0)$
collapses the data for all $\phi$, in agreement with
Eq.~(\ref{pressscaling}). For $\phi > \phi_0$ 
the fluid is no longer the thermodynamically 
stable phase at $T = 0$; these data are shown with open symbols.} 
\end{figure}

The temperature evolution of the equilibrium pressure
obtained in harmonic spheres for densities from $\phi=0.581$ up to 
$\phi=0.846$ is shown in the top panel of
Fig.~\ref{pressure}. For low density, the pressure $Z(\phi,T)$ 
smoothly converges at low temperature 
to the pressure of the equilibrium hard sphere fluid. Indeed we find
that $Z(\phi, T \to 0)$ can be directly compared to the 
direct hard sphere simulations up to $\phi\approx 0.6$. Of course, it 
becomes harder to get to $T \to 0$ at higher density because 
thermalization is too hard to achieve within our computer capabilities. 

To analyze these pressure data, we wish to repeat the analysis 
performed for relaxation times in Sec.~\ref{scaling}.
We have first attempted to collapse the pressure data assuming 
that the equilibrium pressure of the hard sphere fluid 
diverges at point $G$. We thus tried to collapse our data 
using a critical scaling form near $\phi_0$:
\be
Z (\phi,T) \sim f(\phi) H_{\pm}   \left(
\frac{ |\phi_0 - \phi|^{2/\mu} }{T} 
\right),
\ee
which obviously has the same interpretation as in Eq.~(\ref{critical}) 
above.
To this end, we imposed the values for $\mu$ and $\phi_0$ obtained  
in Sec.~\ref{scaling}, and adjusted the function $f(\phi)$ to obtain 
the `best' collapse onto two distinct branches for $\phi$ above
and below $\phi_0$. This approach failed, and we were not able to obtain 
any collapse using this procedure, showing that the scaling properties
of $\tau_\alpha(\phi,T)$ and $Z(\phi,T)$ close to point $G$ are very 
different.

We then made the opposite hypothesis that pressure has a 
smooth behaviour for the fluid of hard spheres approaching 
point $G$ where it stays finite. This hypothesis implies that 
$Z(\phi,T)$ is always a smooth function of temperature.  
In Fig.~\ref{pressure}, we provide evidence supporting 
this hypothesis using the following 
much simpler scaling assumption:
\be
Z(\phi,T) \approx Z(\phi, T \to 0) g(T), 
\label{pressscaling}
\ee 
where $g(T)$ is a simple function of temperature such that 
$g(T\to0)=1$. To produce the collapse in Fig.~\ref{pressure}, we adjusted 
$Z(\phi,T\to0)$ for each volume fraction to get the best collapse of 
the data. By definition, this limit also corresponds to the 
equilibrium pressure of the hard sphere fluid at this volume 
fraction. 

We report the results for $Z(\phi,T\to0)$ in Fig.~\ref{presseq}, where
they can directly be compared with the direct measurements 
performed in Monte Carlo simulations of hard spheres. 
It is obvious that for $\phi \lesssim 0.6$ both data
sets perfectly overlap, confirming the validity of the scaling 
procedure described  in Eq.~(\ref{pressscaling}) in this regime.

Interestingly, the data collapse in Fig.~\ref{pressure} allows 
us to extrapolate the $T \to 0$ behaviour of $Z(\phi,T)$ for volume
fractions at which equilibrium hard sphere simulations are no longer 
available. Therefore, the scaling factor $Z(\phi, T \to 0)$ shown 
in Fig.~\ref{presseq} allows us to extend the fluid equation
of state for hard spheres at larger $\phi$. The
validity of this extrapolation 
simply relies on the reasonable, but probably only 
approximately correct, hypothesis that the temperature 
dependence of $Z(\phi,T)$ does not depend on 
$\phi$ over the limited interval $\phi \in [0.6,0.635]$, which 
is precisely the physical content of Eq.~(\ref{pressscaling}).

Moreover, since the equilibrium relaxation timescales
diverges at $\phi_0$, extrapolation of the fluid branch 
for $\phi>\phi_0$ cannot be performed because a glass 
phase is present. Equilibrium is lost either 
at a finite $T$ as in Fig.~\ref{sketch}-(a), or because the 
$T\to0$ limit is singular, as in Fig.~\ref{sketch}-(b), 
and so the extrapolated value of the fluid
pressure does not coincide with the pressure
of the thermodynamically stable phase above $\phi_0$.

Remarkably, the extension of the fluid branch up to $\phi_0$ shown in 
Fig.~\ref{presseq} continues to follow the BMCSL equation of state. 
This result was not anticipated, since this equation of state is 
constructed to reproduce the behaviour at moderate volume fraction only.
As mentioned above, the BMCSL pressure only diverges at $\phi=1$, 
which is sometimes interpreted as a weakness of this theoretical 
approach because $\phi=1$ is  clearly physically accessible. 
Our results suggest the alternative interesting interpretation that
the BMCSL pressure indeed represents the equation of state of 
the fluid up to large volume fractions, but the fluid is no longer the 
thermodynamically stable phase above $\phi_0$---it is a glass~\cite{FZ}.
Thus, the divergence of the BMCSL fluid equation of state at 
$\phi=1$, and its behaviour above $\phi_0$ are in fact irrelevant.
We shall numerically explore the equation of state 
of the glass phase in the next section.

Finally, the pressure data obtained 
using the scaling behaviour of harmonic 
spheres provide an independent means to estimate the 
pressure of the hard sphere system at the critical volume fraction 
$\phi_0$. The BMCSL equation of state, which represents
the scaled data accurately up to $\phi_0$ hits this critical
density  at the value 
\be
Z (\phi=\phi_0) = 34.4.
\ee 
Obviously, the agreement with the independent estimate 
of $Z_0$ obtained above in Eq.~(\ref{Z0}) is excellent, and 
provides further confidence that the scaling behaviour in 
Eq.~(\ref{pressscaling}) can accurately be used.

To conclude this section, we have 
obtained solid evidence that the equilibrium pressure of 
the fluid of hard spheres is finite at point $G$. 
Thus the phase diagram sketched in Fig.~\ref{sketch}-(b) does not hold, 
and a diverging pressure occuring when all particles are at contact
cannot be reached at thermal equilibrium, suggesting that 
the glass transition at point $G$ does not also 
correspond to a jamming transition. 
In other words, we find that no jamming transition 
can be observed at thermal equilibrium.
In the final section, we shall explore the consequences of this 
finding for understanding the jamming phenomenon.

\section{Exploring multiple glassy states}
\label{jamming}

In this final section, we leave the realm of thermal 
equilibrium to explore glassy states above $\phi_0$,
both for hard and harmonic spheres. As explained 
in Sec.~\ref{model} we shall use hard 
sphere compressions and harmonic sphere annealings 
to reach glassy hard sphere states.
Our two main aims are to investigate the 
equation of state of glasses above $\phi_0$, 
and the possibility to observe jamming transitions
with a diverging pressure, which, we concluded in Sec.~\ref{statics}, 
cannot be explored at thermal equilibrium. 
Since we abandon thermal equilibrium, we must carefully discuss
our numerical protocols, because history now becomes 
part of the story~\cite{comment}.

\subsection{Choice of non-equilibrium protocols}
 
To determine the equilibrium equation of state above $\phi_0$, one 
should in principle get to 
volume fractions $\phi > \phi_0$ while maintaining thermal equilibrium.  
Since the system is non-ergodic, this is not possible in 
computer simulations, but can be done in theoretical 
calculations~\cite{remi}. 

An intuitive numerical solution could be 
to compress a hard sphere system at a finite compression rate, $\Gamma$,  
with the hope that the limit $\Gamma \to 0$ can be reached~\cite{LS1,LS2}.
However, this solution has two immediate drawbacks. First, 
even changing $\Gamma$ by a few orders of magnitude, as can be done 
with present day computers, the system falls out of equilibrium much
above $\phi_0$, so that some extrapolation is again needed~\cite{donev}. 
A second problem stems from the difficulty in such a non-equilibrium path
to check that the system is not undergoing some form of crystallization
(or demixing for a mixture) while being compressed, in which case
the system could end up in configurations that are not necessarily 
representative of the glass states one seeks to investigate.
Since we dedicated much effort to tackle the ordering issue
while studying thermal equilibrium, we must be 
similarly careful when studying glasses.
 
To circumvent the first of these difficulties, we decided 
to present equations of state obtained during compressions without 
attempting any sort of extrapolations. Doing so, we obtain 
pressure measurements at large density that are upper bound to 
the true equilibrium pressure, since non-equilibrium pressure are larger than
the equilibrium ones, just as the energy of an annealed glass 
is larger than the equilibrium energy, recall Eq.~(\ref{dico}). 
Therefore, we will not be able to investigate the nature 
of the thermodynamic transition at $\phi_0$, and the 
possibility for the equilibrium compressibility to have a jump.

To prevent the exploration of
partially ordered or demixed states, we start our compressions (for 
hard spheres)
or annealing (for harmonic spheres) from 
configurations that were produced during our exploration of the 
fluid at thermal equilibrium, for which no tendency to order was detected,
even in very long simulations.  
We then increase the volume fraction with 
a very fast compression rate, 
or decrease the temperature rapidly, 
as described in Sec.~\ref{model}.
We have checked that the particles displacements during these very fast 
compressions are very small, typically much smaller than a particle 
diameter, so that
our final configurations are no more ordered than the 
original fluid states.
Therefore, those states are close in spirit to inherent 
structures usually studied in the context of soft potentials~\cite{SW}.
Interestingly, we find that starting from equilibrium 
configurations, where relaxation was allowed,
the pressure measured during compression or annealing is extremely
weakly dependent on system size. 
We have in fact obtained undistinguishable results for 
$N=1000$ and $N=8000$ particles.
This is in contrast with infinitely fast annealing from
fully random configurations~\cite{pointJ}, which 
are very sensitive to system size.

\subsection{Multiple glasses and jamming densities}

Having fixed compression and annealing rates to very 
large values, we are left with a single control parameter
for exploring glassy states, namely the 
location of the initial equilibrium configuration 
in the $(\phi,T)$ phase diagram, which we now vary.

\begin{figure}
\psfig{file=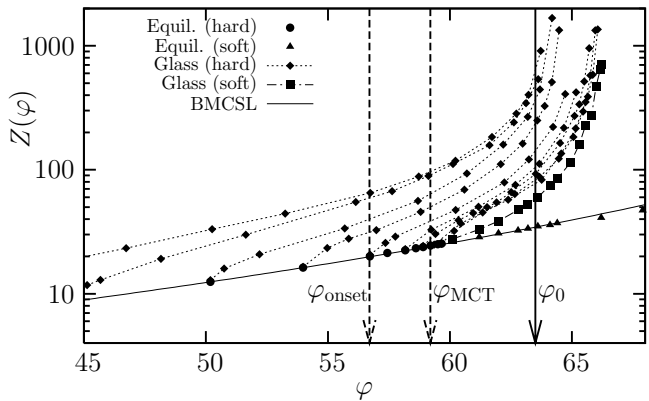,width=8.5cm}
\caption{\label{pasjorge2} 
Various pressure-volume fraction relations obtained from simulations 
of binary hard spheres.
Equilibrium data and BMCSL line are as in Fig.~\ref{presseq}.
The non-equilibrium data from hard sphere compressions and soft 
spheres annealing demonstrate the existence of multiple 
branches, all diverging at different densities, and result
from the glassy behaviour of hard spheres.}
\end{figure} 

We first describe the results obtained during hard sphere 
compressions (along the $T=0$ axis) starting from different initial densities
$\phi_{\rm i}$
in the range  $\phi_{\rm i} \in [0.35, 0.596]$. 
We follow the evolution of the pressure during compression 
using Eq.~(\ref{pressMC}). In Fig.~\ref{pasjorge2} we present 
the results obtained for $N=8000$ particles, averaged over 5 
independent initial configurations, to obtain a better statistics.

As soon as the compression starts, the measured pressure 
deviates from the equilibrium equation of state, emphasizing 
that the compression rate is too fast for the system
to relax to equilibrium, even when volume fraction is not very large.
In the last stages 
of the compression at large volume fraction,
the pressure increases very rapidly with $\phi$, and appears 
to be diverging at some final density, $\phi_{\rm f}$ (note that
Fig.~\ref{pasjorge2} now uses a logarithmic scale for the pressure). 
We stop our compressions when $Z \approx 10^3$, although we could easily 
continue compressing our system up to much larger pressures.

A central observation in Fig.~\ref{pasjorge2} is that the 
compression curves have a strong dependence on $\phi_{\rm i}$, 
which survives even in the $N\to\infty$ limit ($N=1000$ and 8000
already yield consistent results), showing
that the system can be trapped in glassy states that 
explicitely depend on the preparation history.   
Since we have carefully controlled the protocol to avoid mixing 
compressions with partial ordering, these multiple equations
of states result from the non-trivial glassy behaviour
of the hard sphere system, and not from 
a competition between randomness and order.

This discussion equally applies to the terminal densities
$\phi_{\rm f}$ of these compression branches where
pressure diverges. 
These configurations are the hard sphere analogs
of the inherent structures obtained at $T=0$ in systems with 
soft potentials, and $1/\phi_{\rm f}$ plays a role similar to the
potential energy of inherent structures. The fact that $\phi_{\rm f}$
has a strong dependence on $\phi_{\rm i}$ thus corresponds to 
the observation that the energy of inherent structures decreases 
when the temperature is decreased~\cite{sri1}, 
as already discussed in Refs.~\cite{krauth,dave} 
in the context of hard spheres.  
We find that $\phi_{\rm f}$ increases slowly from $\approx 0.642$ 
when $\phi_{\rm i}$ is in the dilute 
regime, but it starts increasing more rapidly when 
$\phi_{\rm i}$ gets larger than some `onset' volume fraction,
$\phi_{\rm onset} \approx 0.56$, which marks the 
onset of slow dynamics for the hard sphere fluid~\cite{dave}. 
It then grows 
markedly up to $\phi_{\rm f} \approx 0.662$ for the largest 
$\phi_{\rm i}$ considered in this work. 
Thus our results suggest that glassiness 
alone can be responsible for the existence of a finite range 
of volume fractions where hard sphere 
configurations get jammed, and give an alternative explanation
as to why the concept of random close packing is not 
`well-defined', 
for which the existence of the crystalline phase is 
irrelevant~\cite{torquato}.

A final conclusion drawn from the compression curves 
in Fig.~\ref{pasjorge2} is that 
the pressure at $\phi_0$ for non-equilibrium 
compressions is finite. Since these
non-equilibrium measurements represent upper
bounds to the equilibrium pressure, we obtain a direct 
verification that 
the equilibrium pressure of the fluid does not diverge at 
$\phi_0$. This observation does not involve fits or extrapolations.

Turning finally to harmonic spheres, we have found very similar results. 
We find that the $T \to 0$ limit of the pressure obtained in fast annealing
of equilibrated configurations at finite temperature also strongly 
depends on the initial temperature, in full agreement with
studies of inherent structures in systems with soft potentials~\cite{sri1}.  
In Fig.~\ref{pasjorge2}, we present the result for the $T \to 0$ 
pressure obtained when choosing, for each volume fraction, 
an initial configuration corresponding to the lowest temperature
for which thermal equilibrium had been reached (see the phase diagram
in Fig.~\ref{phaselong}). The glass equation of state obtained this
way is qualitatively very similar to the one obtained by fast 
compressions of well equilibrated hard spheres, and it diverges 
near $\phi_{\rm f} \approx 0.664$. 

This rather large volume fraction, $\phi_{\rm f} = 0.664$,
provides two interesting perspectives. First, if one insists 
that thermal equilibrium can be maintained as
long as pressure is finite, then one is led to conclude that 
a glass transition 
with a diverging relaxation time can only occur 
at least above $\phi=0.664$. However, when analyzing our equilibrium dynamic
data in Sec.~\ref{dynamics}, we were never able to obtain 
such a large value for the critical density $\phi_0$.
Second, if we use the theoretical perspective
sketched in Fig.~\ref{sketch}-(a) with the existence 
of a true glass phase for hard spheres,  
what we have obtained here is a lower bound on the location of the 
Glass Close Packing density, $0.664 < \phi_{GCP}$.

\section{Conclusion}
\label{conclusion}

In this section, we summarize and discuss the main
results of our numerical study. 

\subsection{Summary}

We studied a binary mixture of hard spheres
at increasing volume fractions. We found that 
dynamics slows down dramatically, and were able to follow 
the first 7 decades of relaxation times. We showed that 
an algebraic power law, as predicted by mode-coupling theory,
only describes a window of about 3 decades immediately after the 
onset of glassy dynamics near $\phi_{\rm onset} \approx 0.56$.
The dynamics is best fitted with a generalized VFT law, 
Eq.~(\ref{activatedHS}), with the best fit obtained for $\phi_0 \approx 
0.635$ and $\delta \approx 2.2$, but fits with the traditional 
value $\delta=1$ diverging near $\phi_{\rm VFT} \approx 0.615$ 
were acceptable. 

We found that the volume fraction dependence of the equilibrium pressure 
was rather modest, and were not able to extrapolate these data
to obtain a critical volume fraction where the pressure diverges, 
implying that $\tau_\alpha$ and $Z$ do not appear to diverge
at the same volume fraction, in contrast with free volume 
arguments. 

We suggested to build an `Angell plot for hard spheres', 
representing the evolution of $\log \tau_\alpha$ vs. $Z$, parametrized
by $\phi$. In this representation, the density analog of the 
Arrhenius behaviour,  Eq.~(\ref{arrZ}), appears as a straight line. 
We find instead that hard spheres display a `fragile' behaviour, 
since the pressure dependence of $\tau_\alpha$ is much more marked.   
In this representation, we could show that several 
theoretical predictions for the dynamic behaviour of hard spheres
do not describe our data, the best description being 
offered by the hypothesis that dynamics diverges at a finite 
pressure. We estimated $Z_0 \approx 34.4$ for the present system. 

We have shown that extending these study  
using temperature as a second, independent
control parameter, could provide an independent 
confirmation of these conclusions 
about the dynamic and thermodynamic behaviour of hard spheres. 
These temperature studies were moreover able to 
make the above statements quantitatively much more convincing. 
By using dynamic scaling ideas, we determined the location $\phi_0$
of the divergence, and the functional form of the 
relaxation time with a better precision, showing in particular
that the most commonly used value $\delta=1$ in Eq.~(\ref{activatedHS}) 
was incompatible with our data. For thermodynamic quantities, 
we discovered a much simpler scaling behaviour of the 
pressure, allowing to confirm that $Z$ is not divergent at 
the dynamic transition $\phi_0$.   

We have directly confirmed this statement by 
devising non-equilibrium paths to explore glassy states 
at large volume fractions, carefully dealing with 
crystallization and demixing issues. We have found 
a family of non-equilibrium equations of state 
in the glass phase, which can be seen as upper bounds
to the true equilibrium equation of state above $\phi_0$.    
We have shown that these multiple branches
remain distinct in the thermodynamic limit, and diverge 
in a broad range of volume fractions between $\phi=0.642$ and 
$\phi=0.664$, 
where configurations get jammed. 

\subsection{Discussion and open questions}

Gathering the above results sheds light on the possible 
phase diagrams for the harmonic sphere system sketched 
in Fig.~\ref{sketch}, and its $T=0$ limit where it coincides 
with the hard sphere system. 

The best description of our data is obtained 
if we assume that the relaxation time in the hard sphere 
limit diverges at $\phi_0 \approx 0.635$, where the 
equilibrium pressure of the fluid is $Z_0 \approx 34.4$. 
Although these numerical values are specific to the present
binary mixture studied in this work,  
these results invalidate a number of theoretical approaches
predicting $Z_0 = \infty$, 
and more generally the idea that the dynamic arrest 
in hard spheres simply occurs when the interparticle distance 
vanishes~\cite{freevol,liu,ohern,ken}. The only theoretical scenario where a 
glass transition occurs at finite pressure is the one 
predicted by theories and calculations 
based on the concept of an ideal glass transition 
of the random first order type~\cite{KTW,FZ,FZ2}. 
So, we are led to the conclusion 
that our data provide strong support for this scenario in the 
context of hard spheres.  We were not able to 
provide a similar evidence for the temperature
behaviour of harmonic spheres above $\phi_0$, and the 
possibility of a finite temperature glass line as in Fig.~\ref{sketch}-(a)
remains an open issue. 

Since we opened the paper with ironic remarks about such sharp claims
about the existence of glass transitions, let us make here a series of 
cautious remarks about the above statement for hard spheres.
First, we repeat that our conclusions 
are based upon solid, but necessarily limited in range, numerical evidence.
Thus, we leave open the possiblity that the picture changes when 
a broader range of timescales is covered. But we also insist that the 
range covered numerically is just as large as the range covered 
experimentally in colloids~\cite{luca}, while work in the field 
of molecular glasses suggest that the physics hardly changes
when several decades of relaxation timescales are added.
Thus, we can at least claim experimental relevance for our 
results.

Second, although we suggest that the nature of point $G$ is consistent
with an ideal glass transition of the random first order type, 
its precise nature remains to be established. Theoretically,
such a transition is defined by the vanishing of the configurational
entropy counting the number of metastable states. Thus, 
it would be important to measure 
the configurational entropy numerically in the vicinity of 
point $G$. Unfortunately, only approximations 
to the configurational entropy can be accessed 
numerically, because the very concept of metastable states 
is not well-defined~\cite{jorge3}.  
Another potential problematic aspect concerns the dynamical 
behaviour predicted within random first order theory: the exponent
$\delta \approx 2.2$ in Eq.~(\ref{activatedHS}) is usually not the one
used when analyzing experimental data in molecular glasses, and is 
not the one predicted in Ref.~\cite{KTW}, although 
scaling arguments~\cite{BB} suggest ways out of the problem that
remain to be worked out. 
Also, since explicit replica calculations only exist in the 
hard sphere limit, it would certainly be worthwile 
to extend the computation to harmonic spheres at finite temperature.
Work is in progress in this direction.
 
The third remark of caution stems from the existing
line of research which aims at demonstrating that an ideal
glass transition cannot exist in hard spheres. 
In Sec.~\ref{rcp}, we discussed why we believe published 
theoretical arguments do not establish the irrelevance of a concept 
of an ideal glass transition in hard spheres, even for the specific case 
considered in Ref.~\cite{mixture}. 
Numerical evidence against the existence 
of an ideal glass transition was also given in Ref.~\cite{krauth}. 
We stress, 
however, that the location of $\phi_0$ in this last work was defined
using algebraic laws for the relaxation time, as in Eq.~(\ref{mctPHI}). 
Using a similar
description we would have (incorrectly) located $\phi_0$ close 
to $\phi_{\rm MCT} \approx 0.592$, where we indeed were able to 
show that the pressure is not singular, in agreement with Ref.~\cite{krauth}. 
Therefore,
the path  to disprove our conclusions about the nature 
and location of point $G$ is clear: 
one should provide numerical evidence that thermal equilibrium 
can be maintained for the present 
binary mixture with no pressure singularity for 
volume fractions larger than $\phi_0 \approx 0.635$ and pressures 
larger than $Z_0 \approx 34.4$. We believe this is a hard numerical
task, even when using smart algorithms~\cite{dave2}.

A final suggestion for future research is the connection 
to jamming transitions suggested in Fig.~\ref{pasjorge2}.
We provided direct evidence that a `random close packing'
density can not be provided in a unique manner, as 
jamming transitions can occur within a finite range of 
volume fractions. Our argument is different from 
the one presented in Ref.~\cite{torquato}, since it does not rely
on the existence of a crystalline or ordered phase. 
It is, however, in good agreement with recent work using 
ideas from random first order theory to study 
jamming of hard spheres~\cite{jorge1,jorge2,FZ}. 
In future work, we shall 
study more precisely the final configurations 
obtained when pressure diverges in Fig.~\ref{pasjorge2}.
Our conclusion that, along the metastable fluid 
branch, an ideal glass transition intervenes 
at finite pressure shows that a `reproducible' 
jamming transition cannot be observed at thermal 
equilibrium. This shows also that the glass transition 
is likely not driven by geometric jamming, but we can 
tentatively use the jamming phase diagram of Ref.~\cite{liunagel}
in the opposite direction and claim that glassiness
observed in glass-forming liquids carries indeed 
interesting consequences for understanding jamming transitions. 

\begin{acknowledgments}
We thank G. Biroli, P. Chaudhuri, 
L. Cipelletti, R. Jack, H. Jacquin, J. Kurchan, 
G. Tarjus, and F. Zamponi for useful discussions 
and criticisms, and Alela Diane and Patricia
Barber for the rhymes accompanying completion
of a manuscript ``composed for connoisseurs, 
for the refreshment of their spirits~\cite{bach}''.
\end{acknowledgments}

\end{document}